\documentclass[11pt]{article}

\usepackage{fullpage}
\usepackage{latexsym}
\usepackage{amsfonts}
\usepackage{amsmath}

\newcommand{\ket}[1]{|#1\rangle}
\newcommand{\bra}[1]{\langle#1|}
\newcommand{\inp}[2]{\langle{#1}|{#2}\rangle}

\def\01{\{0,1\}}
\def\e{\varepsilon}

\newcommand{\cSWAP}{\mbox{\rm c-SWAP}}

\begin{document}

\title{\Large\bf Quantum fingerprinting}

\author{Harry Buhrman\thanks{CWI, P.O.~Box 94709, Amsterdam, The Netherlands.
Email: $\{${\tt buhrman},{\tt rdewolf}$\}${\tt @cwi.nl}.
Partially supported by the EU fifth framework project QAIP, IST--1999--11234.}
\and 
Richard Cleve\thanks{Department of Computer Science, University of Calgary, 
Calgary, Alberta, Canada T2N 1N4.
Email: $\{${\tt cleve},{\tt jwatrous}$\}${\tt @cpsc.ucalgary.ca}.
Partially supported by Canada's NSERC.}
\and 
John Watrous$^{\dag}$
\and 
Ronald de Wolf$^{\ast}$
}

\date{\empty}

\maketitle

\begin{abstract}
Classical fingerprinting associates with each string a shorter string 
(its {\em fingerprint}), such that, with high probability, any two distinct 
strings can be distinguished by comparing their fingerprints alone.
The fingerprints can be exponentially smaller than the original strings 
if the parties preparing the fingerprints share a random key, but not if they
only have access to uncorrelated random sources.
In this paper we show that fingerprints consisting of {\em quantum\/} 
information {\em can} be made exponentially smaller than the original strings 
without any correlations or entanglement between the parties: we give a 
scheme where the quantum fingerprints are exponentially shorter than the 
original strings and we give a test that distinguishes any two unknown quantum 
fingerprints with high probability.
Our scheme implies an exponential quantum/classical gap for the equality 
problem in the simultaneous message passing model of communication complexity.
We optimize several aspects of our scheme.
\end{abstract}


\section{Introduction}\label{secintro}

Fingerprinting can be a useful mechanism for determining if two strings 
are the same: each string is associated with a much shorter fingerprint 
and comparisons between strings are made in terms of their fingerprints 
alone.
This can lead to savings in the communication and storage of information.

The notion of fingerprinting arises naturally in the setting of 
{\em communication complexity\/} (see \cite{kushilevitz&nisan:cc}).
The particular model of communication complexity that we consider in this 
paper is called the {\em simultaneous message passing\/} model, which was 
introduced by Yao~\cite{yao:distributive} in his original paper on 
communication complexity.
In this model, two parties---Alice and Bob---receive inputs $x$ and $y$, 
respectively, and are not permitted to communicate with one another directly.
Rather they each send a message to a third party, called the {\em referee}, 
who determines the output of the protocol based solely on the messages 
sent by Alice and Bob.
The collective goal of the three parties is to cause the protocol to 
output the correct value of some function $f(x,y)$ while minimizing the 
amount of information that Alice and Bob send to the referee.

For the {\em equality} problem, the function is simply
\[
f(x,y) = \left\{\begin{array}{ll}1&\mbox{if $x=y$}\\0&\mbox{if $x\not=y$.}
\end{array}\right.
\]
The problem can of course be trivially solved if Alice sends $x$ and Bob 
sends $y$ to the referee, who can then simply compute $f(x,y)$.
However, the cost of this protocol is high; if $x$ and $y$ are $n$-bit 
strings, then a total of $2n$ bits are communicated.
If Alice and Bob instead send {\em fingerprints} of $x$ and $y$,
which may each be considerably shorter than $x$ and $y$, the cost can be
reduced significantly.
The question we are interested in is how much the size of the fingerprints
can be reduced.

If Alice and Bob share a random $O(\log n)$-bit key then the fingerprints 
need only be of {\em constant\/} length if we allow a small probability 
of error; a brief sketch of this follows.
A binary error-correcting code is used, which can be represented as 
a function $E : \01^n \rightarrow \01^m$, where $E(x)$ is the 
codeword associated with $x \in \01^n$.
There exist error-correcting codes (Justesen codes, for instance) with $m = cn$
such that the Hamming distance between any two distinct codewords $E(x)$ and
$E(y)$ (with $x \neq y$) is at least $(1 - \delta)m$, where $c$ and 
$\delta$ are constants.
For the particular case of Justesen codes, we may choose any $c>2$ and we will
have $\delta< 9/10 + 1/(15c)$ (assuming $n$ is sufficiently large).
For further information on Justesen codes, see Justesen~\cite{justesen}
and MacWilliams and Sloane~\cite[Chapter 10]{MacWilliamsS77}.
Now, for $x \in \01^n$ and $i \in \{1,2,\ldots,m\}$, let $E_i(x)$ denote 
the $i^{\mbox{\scriptsize th}}$ bit of $E(x)$.
The shared key is a random $i \in \{1,2,\ldots,m\}$ (which consists 
of $\log(m) \in \log(n) + O(1)$ bits).
Alice and Bob respectively send the bits $E_i(x)$ and $E_i(y)$ to the 
referee, who then outputs 1 if and only if $E_i(x) = E_i(y)$.
If $x=y$ then $E_i(x) = E_i(y)$, so then the outcome is correct.
If $x \neq y$ then the probability that $E_i(x) = E_i(y)$ is at most 
$\delta$, so the outcome is correct with probability $1 - \delta$.
The error probability can be reduced from $\delta$ to any $\e > 0$ by 
having Alice and Bob send $O(\log(1/\e))$ independent random bits of 
the codewords $E(x)$ and $E(y)$ to the referee.
In this case, the length of each fingerprint is $O(\log(1/\e))$ bits.

One disadvantage of the above scheme is that it requires overhead in 
creating and maintaining a shared key.
Moreover, once the key is distributed, it must be stored securely until 
the inputs are obtained.
This is because an adversary who knows the value of the key can easily 
choose inputs $x$ and $y$ such that $x\neq y$ but for which the output 
of the protocol always indicates that $x = y$.

Yao~\cite[Section~4.D]{yao:distributive} posed as an open problem the
question of what happens in this model if Alice and Bob do not have a shared
key.
Ambainis~\cite{ambainis:3computer} proved that fingerprints of $O(\sqrt{n})$
bits suffice if we allow a small error probability (see also 
\cite{knr:rand1round,newman&szegedy:1round,babai&kimmel:simultaneous}).
Note that in this setting Alice and Bob still have access to random bits, but
their random bits may not be correlated.
Subsequently, Newman and Szegedy~\cite{newman&szegedy:1round} proved
a matching lower bound of $\Omega(\sqrt{n})$.
Their result was generalized by
Babai and Kimmel~\cite{babai&kimmel:simultaneous} to the result that the
randomized and deterministic complexity can be at most quadratically far
apart for {\em any} function in this model.
Babai and Kimmel attribute a simplified proof of this fact to Bourgain and
Wigderson.

We shall consider the problem where there is no shared key (or entanglement)
between Alice and Bob, but the fingerprints can consist of quantum information.
In Section~\ref{secfingerprint}, we show that $O(\log n)$-qubit fingerprints 
are sufficient to solve the equality problem in this setting---an exponential 
improvement over the $\sqrt{n}$-bound for the comparable classical case.
Our method is to set the $2^n$ fingerprints to quantum states 
whose pairwise inner-products are bounded below 1 in absolute
value and to use a test that identifies identical fingerprints and
distinguishes distinct fingerprints with good probability.
(It is possible to take the fingerprints to be nearly pairwise orthogonal, 
although the bound on the absolute value of the inner product between pairs 
of states is not directly related to the error probability of the 
fingerprinting method.)
This gives a simultaneous message passing protocol for equality in the 
obvious way: Alice and Bob send the fingerprints of their respective inputs 
to the referee, who then executes the test to check if the fingerprints 
are equal or distinct.
In Section~\ref{secfingerprint}, we also show that the fingerprints must 
consist of at least $\Omega(\log n)$ qubits if the error probability 
is bounded below 1.

In Sections \ref{secimprovedfp} and \ref{secdistinguish}, we consider 
possible improvements to the efficiency of the fingerprinting methods 
of Section~\ref{secfingerprint}.
In Section~\ref{secimprovedfp}, we investigate the number of qubits 
required to contain $2^n$ fingerprints with pairwise inner product 
bounded in absolute value by any $\delta < 1$.
In Section~\ref{secdistinguish}, we consider the efficiency of tests 
that distinguish between $k$ copies of pairs of indentical states 
and $k$ copies of pairs of states whose inner product is bounded in 
absolute value by any $\delta < 1$.

Finally, in Section~\ref{secexactfinger} we consider a variation of
fingerprinting with a shared {\em quantum} key, consisting of $O(\log n)$
shared Bell states (EPR-pairs).
We observe that results in~\cite{bct:simulating} imply that
errorless (i.e., exact) fingerprinting is possible with $O(\log n)$-bit 
classical fingerprints in a particular context where achieving the same 
performance with only a classical shared key requires fingerprints of length 
$\Omega(n)$.

We assume the reader is familiar with the basic notions of quantum
computation and quantum information---for further information we refer the
reader to the book by Nielsen and Chuang~\cite{nielsen&chuang:qc}.


\section{Quantum fingerprinting without shared keys}\label{secfingerprint}

In this section, we show how to solve the equality problem in the simultaneous 
message passing model with logarithmic-length quantum fingerprints in a 
context where no shared key is available.
The solution is quite simple and the fingerprints are exponentially 
shorter than in the comparable classical setting, where $\Theta(\sqrt{n})$ 
bit fingerprints are necessary and sufficient 
(see the references in the introduction).
The method that we present is based on classical error-correcting codes,
though in a different manner than discussed in Section~\ref{secintro}
since no shared key is available.

Assume that for fixed $c>1$ and $\delta<1$ we have an error correcting code
$E: \01^n \rightarrow \01^m$ for each $n$, where $m = cn$ and such that the
distance between distinct codewords $E(x)$ and $E(y)$ is at least
$(1-\delta)m$.
As mentioned in Section~\ref{secintro}, a reasonable first choice of 
such codes are Justesen codes, which give $\delta < 9/10 + 1/(15c)$
for any chosen $c>2$.
Now, for any choice of $n$, we define the $(\log(m)+1)$-qubit state
$\ket{h_x}$ as
\begin{equation}\label{eqhx}
\ket{h_x} = {\textstyle{\frac{1}{\sqrt{m}}}}
\sum_{i = 1}^m \ket{i}\ket{E_i(x)}
\end{equation}
for each $x \in \01^n$.
Since two distinct codewords can be equal in at most $\delta m$ positions, 
for any $x\neq y$ we have $\inp{h_x}{h_y} \le \delta m / m = \delta$.
Thus we have $2^n$ different $(\log(n)+O(1))$-qubit states, and each pair 
of them has inner product at most $\delta$.

The simultaneous message passing protocol for the equality problem 
works as follows.
When given $n$-bit inputs $x$ and $y$, respectively, Alice and Bob 
send fingerprints $\ket{h_x}$ and $\ket{h_y}$ to the referee.
%
%
%
Then the referee must distinguish between the case where the two 
states received---call them $\ket{\phi}$ and $\ket{\psi}$---are 
identical or have inner product at most $\delta$.
This is accomplished with one-sided error probability by the 
procedure that measures and outputs the first qubit of the state
\[
(H \otimes I)(\cSWAP)(H \otimes I)\ket{0}\ket{\phi}\ket{\psi}.
\]
Here $H$ is the Hadamard transform, which maps
$\ket{b}\rightarrow\frac{1}{\sqrt{2}}(\ket{0}+(-1)^b\ket{1})$, 
SWAP is the operation $\ket{\phi}\ket{\psi}\rightarrow\ket{\psi}\ket{\phi}$
and c-SWAP is the controlled-SWAP (controlled by the first qubit).
The circuit for this procedure is illustrated in Figure~\ref{figdist}.
\begin{figure}[hbt]
\centering
\setlength{\unitlength}{0.3mm}
\begin{picture}(200,120)
\put(-10,80){\makebox(20,20){$\ket{0}$}}
\put(-10,40){\makebox(20,20){$\ket{\phi}$}}
\put(-10,0){\makebox(20,20){$\ket{\psi}$}}
\put(210,80){\makebox(20,20){\small measure}}
\put(10,90){\line(1,0){30}}
\put(40,80){\framebox(20,20){$H$}}
\put(140,80){\framebox(20,20){$H$}}
\put(60,90){\line(1,0){80}}
\put(160,90){\line(1,0){30}}
\put(100,90){\line(0,-1){30}}
\put(100,90){\circle*{5}}
\put(80,0){\framebox(40,60){\small SWAP}}
\put(10,50){\line(1,0){70}}
\put(10,10){\line(1,0){70}}
\put(120,50){\line(1,0){70}}
\put(120,10){\line(1,0){70}}
\end{picture}
\caption{Circuit to test if $\ket{\phi}=\ket{\psi}$ or 
$|\inp{\phi}{\psi}| \le \delta$}\label{figdist}
\end{figure}
By tracing through the execution of this circuit, one can determine that 
the final state before the measurement is 
\[
{\textstyle \frac{1}{2}\ket{0}(\ket{\phi}\ket{\psi}+\ket{\psi}\ket{\phi}) + 
\frac{1}{2}\ket{1}(\ket{\phi}\ket{\psi}-\ket{\psi}\ket{\phi})}.
\]
Measuring the first qubit of this state produces outcome 1 with probability 
$\frac{1}{2} - \frac{1}{2}|\inp{\phi}{\psi}|^2$.
This probability is 0 if $x = y$ and is at least $\frac{1}{2}(1-\delta^2)>0$ 
if $x \neq y$.
Thus, the test determines which case holds with one-sided error
$\frac{1}{2}(1 + \delta^2)$.

The error probability of the test can be reduced to any $\e > 0$ by setting 
the fingerprint of $x \in \01^n$ to $\ket{h_x}^{\otimes k}$ 
for a suitable $k \in O(\log(1/\e))$.
{}From such fingerprints, the referee can independently perform the test in 
Figure~\ref{figdist} $k$ times, resulting in an error probability 
below $\e$.
In this case, the length of each fingerprint is $O((\log n)(\log(1/\e))$.

It is worth considering what goes wrong if one tries to simulate the 
above quantum protocol using classical mixtures in place of quantum 
superpositions.
In such a protocol, Alice and Bob send $(i,E_i(x))$ and $(j,E_j(y))$ 
respectively to the referee for {\em independent\/} random uniformly 
distributed $i,j \in \{1,2,\ldots,m\}$.
If it should happen that $i = j$ then the referee can make a 
statistical inference about whether or not $x = y$.
But $i = j$ occurs with probability only $O(1 / n)$---and the 
ability of the referee to make an inference when $i \neq j$ 
seems difficult.
For many error-correcting codes, no inference whatsoever about $x=y$ is 
possible when $i \neq j$ and the lower bound in \cite{newman&szegedy:1round} 
implies that no error-correcting code enables inferences to be made when 
$i \neq j$ with error probability bounded below 1.
The distinguishing test in Figure~\ref{figdist} can be viewed as 
a quantum operation which has no analogous classical probabilistic 
counterpart.

Our quantum protocol for equality in the simultaneous message model 
uses $O(\log n)$-qubit fingerprints for any constant error probability.
Is it possible to use fewer qubits?
In fact, without a shared key, $\Omega(\log n)$-qubit fingerprints 
are necessary.
This is because any $k$-qubit quantum state can be specified 
within exponential precision with $O(k 2^k)$ classical bits.
Therefore the existence of a $k$-qubit quantum protocol implies the 
existence of an $O(k 2^k)$-bit (deterministic) classical protocol.
{}From this we can infer that $k \in \Omega(\log n)$.


\section{Sets of pairwise-distinguishable states in
low-dimensional spaces}\label{secimprovedfp}

In Section~\ref{secfingerprint}, we employed a particular classical 
error-correcting code to construct a set of $2^n$ quantum states 
with pairwise inner products below $\delta$ in absolute value.
Here, we consider the question of how few qubits are sufficient 
for this to be accomplished for an arbitrarily small $\delta > 0$.
We show that $\log n + O(\log(1/\delta))$ qubits are sufficient.
While this gives somewhat better bounds than the Justesen codes discussed
in Section~\ref{secfingerprint}, unfortunately we only have a nonconstructive
proof of this fact.
The proof follows.

Suppose $d\geq\frac{4n}{\delta^2\log e}$.
Then we claim there are $2^n$ unit vectors in $\mathbb{R}^d$ with pairwise
inner product at most $\delta$ in absolute value.
Consider two random vectors in $v,w$ in $\{+1,-1\}^d/\sqrt{d}$.
Suppose $v$ and $w$ agree in $d'$ coordinates and disagree in $d-d'$ 
coordinates, then their inner product is $\inp{v}{w}=(2d'-d)/d$.
Using a Chernoff bound~\cite[Corollary~A.2]{alon&spencer:probmethod} we have
$$
\Pr[|\inp{v}{w}|>\delta]=\Pr[|2d'-d|>\delta d]\leq 2e^{-\delta^2 d/2}.
$$
Now pick a set $S$ of $2^n$ random vectors from $\{+1,-1\}^d/\sqrt{d}$.
The probability that there are distinct $v,w\in S$ with
large inner product is upper bounded by
\begin{eqnarray*}
\Pr[\exists \mbox{ distinct }v,w\in S \mbox{ with } |\inp{v}{w}|>\delta]
& \leq &
\sum_{\mbox{\tiny distinct }v,w \in S}\Pr[|\inp{v}{w}|>\delta]\\
& < & \binom{2^n}{2}2e^{-\delta^2 d/2}
< 2^{2n-\delta^2d\log e/2}.
\end{eqnarray*}
If $d\geq 4n/\delta^2\log e$ then this probability is $<1$, which implies the 
existence of a set $S$ of $2^n$ vectors having the right properties.

By associating $\01^n$ with the $2^n$ vectors above, we obtain fingerprints 
of $\log(4n/\delta^2\log e) \in \log n + O(\log(1/\delta))$ qubits for any 
$\delta > 0$.

Up to constant factors, the nonconstructive method above is optimal in the 
following sense.
Let $\delta \geq 2^{-n}$.
Then an assignment of $b$-qubit states to all $n$-bit strings
such that the absolute value of the inner product between any two fingerprints
is at most $\delta$, requires $b\in \Omega(\log(n/\delta))$ qubits.
In order to demonstrate this, we will prove and then combine two lower bounds
on $b$.

Firstly, the states can be used as fingerprints to solve the equality
problem of communication complexity with bounded-error probability
in one round of communication (Alice sends the fingerprint of her 
input $x$ to Bob, who compares it with the fingerprint of his $y$).
Therefore the known lower bound for equality 
implies $b\geq c\log n$ for some $c>0$.

Secondly, pick a set of $a=1/\delta$ different fingerprints.
These are complex unit vectors $v_1,\ldots,v_a$ of dimension $2^b$, 
whose pairwise inner products are at most $\delta$ in absolute value.
Let $A$ be the $a\times 2^b$ matrix having the conjugated vectors 
$v_i$ as rows and let $B$ be the $2^b\times a$ matrix having the $v_i$
as columns. Consider the $a\times a$ matrix $C=AB$. Its $i,j$ entry is
$C_{ij}=\inp{v_i}{v_j}$, so the diagonal entries of $C$ are 1, 
the off-diagonal entries are at most $\delta$ in absolute value. 
This means that $C$ is strictly diagonally dominant: 
$C_{ii}=1>(a-1)\delta\geq\sum_{j\neq i}|C_{ij}|$ for all $i$.
It is known that such a matrix has full rank
\cite[Theorem~6.1.10.a]{horn&johnson:ma}.
This implies that the $a$ vectors $v_1,\ldots,v_a$ are linearly 
independent and hence must have dimension at least $a$.
Thus $1/\delta=a\leq 2^b$, hence $b\geq\log(1/\delta)$.

Since both lower bounds on $b$ hold simultaneously, we have
$$
b\geq\max\{c\log n, \log(1/\delta)\}\geq \frac{c\log n+\log(1/\delta)}{2}
\in\Omega(\log(n/\delta)).
$$

It should be noted that having small inner product $\delta$ is desirable but
not all-important.
For instance, there is a trade-off between $\delta$ and the number of copies
of each state sent by Alice and Bob in the simultaneous message passing
protocol for equality from the previous section in terms of the total number
of qubits communicated and the resulting error bound.


\section{The state distinguishing problem}\label{secdistinguish}

Motivated by the fingerprinting scheme of Section~\ref{secfingerprint}, 
we define the {\em state distinguishing problem\/} as follows.
The input consists of $k$ copies of each of two quantum states $\ket{\phi}$
and $\ket{\psi}$ with a promise that the two states are either identical
or have inner product bounded in absolute value by some given $\delta<1$.
The goal is to distinguish between the two cases with as high probability
as possible.

One method for solving this problem is to use the method in 
Section~\ref{secfingerprint}, independently performing the test in 
Figure~\ref{figdist} $k$ times, resulting in an error probability of 0 
in the identical case and $(\frac{1+\delta^2}{2})^{k}$ otherwise.
We will describe an improved method, whose error probability is 
approximately $\sqrt{\pi k}(\frac{1+\delta}{2})^{2k}$ (which is almost 
a quadratic reduction when $\delta$ is small).
We also show that this is nearly optimal by proving a lower bound of 
$\frac{1}{4}(\frac{1+\delta}{2})^{2k}$ on the error probability.

The improved method for the state distinguishing problem uses registers 
$R_1,\ldots,R_{2k}$, which initially contain $\ket{\phi},\ldots,\ket{\phi},
\ket{\psi},\ldots,\ket{\psi}$ ($k$ copies of each).
It also uses a register $P$ whose classical states include encodings of 
all the permutations in $S_{2k}$.
Let $0$ denote the identity permutation and let $P$ be initialized to $0$.
Let $F$ be any transformation satisfying
\[
F:\ket{0}
\mapsto\frac{1}{\sqrt{(2k)!}}\sum_{\sigma\in S_{2k}}\ket{\sigma}.
\]
Such a transformation can easily be computed in polynomial time.

The distinguishing procedure operates as follows:

\begin{enumerate}
\item
Apply $F$ to register $P$.
\item
Apply a conditional permutation on the contents of registers 
$R_1,\ldots,R_{2k}$, conditioned on the permutation specified in $P$.
\item
Apply $F^{\dagger}$ to $P$ and measure the final state.
If $P$ contains $0$ then answer {\em equal}, otherwise answer 
{\em not equal}.
\end{enumerate}
The state after is Step~2 is 
\[
\frac{1}{\sqrt{(2k)!}}\sum_{\sigma\in S_{2k}}\ket{\sigma}
\sigma(\ket{\phi}\cdots\ket{\phi}\ket{\psi}\cdots\ket{\psi})
\]
(where $\sigma(\ket{\phi}\cdots\ket{\phi}\ket{\psi}\cdots\ket{\psi})$
means we permute the contents of the $2k$ registers according to 
$\sigma$).\vspace{2mm}

\noindent{\bf Case 1:} $\ket{\phi}=\ket{\psi}$.
In this case the permutation of the registers does absolutely
nothing, so the procedure answers {\em equal} with certainty.

\vspace{2mm}
\noindent{\bf Case 2:} Assume $|\langle\phi|\psi\rangle|<\delta$.
The probability of answering {\em equal} is the squared norm of the vector
obtained by applying the projection $\ket{0}\bra{0}\otimes I$ to the final
state, which is
\begin{eqnarray*}
p_{eq} & = &
\left\|\frac{1}{\sqrt{(2k)!}}\sum_{\sigma\in S_{2k}}
\langle 0|F^{\dagger}|\sigma\rangle
\sigma(\ket{\phi}\cdots\ket{\phi}\ket{\psi}\cdots\ket{\psi})\right\|^2\\
& = &
\left\|\frac{1}{(2k)!}\sum_{\sigma\in S_{2k}}
\sigma(\ket{\phi}\cdots\ket{\phi}\ket{\psi}\cdots\ket{\psi})\right\|^2.
\end{eqnarray*}
Since $\|\ket{\eta}\|^2 = \langle \eta|\eta\rangle$ for any $\ket{\eta}$
we may simplify this probability as follows:
\begin{eqnarray*}
p_{eq} & = &
\frac{1}{((2k)!)^2}\sum_{\sigma,\tau\in S_{2k}}
\sigma(\bra{\phi}\cdots\bra{\phi}\bra{\psi}\cdots\bra{\psi})
\tau(\ket{\phi}\cdots\ket{\phi}\ket{\psi}\cdots\ket{\psi})\\
& = &
\frac{1}{((2k)!)^2}\sum_{\sigma,\tau\in S_{2k}}
\bra{\phi}\cdots\bra{\phi}\bra{\psi}\cdots\bra{\psi}
\sigma^{-1}\tau(\ket{\phi}\cdots\ket{\phi}\ket{\psi}\cdots\ket{\psi})\\
& = &
\frac{1}{(2k)!}\sum_{\sigma\in S_{2k}}
\bra{\phi}\cdots\bra{\phi}\bra{\psi}\cdots\bra{\psi}
\sigma(\ket{\phi}\cdots\ket{\phi}\ket{\psi}\cdots\ket{\psi})\\
& = &
\frac{(k!)^2}{(2k)!}\sum_{j=0}^k\binom{k}{j}^2\delta^{2j}\\
& \le &
\frac{(k!)^2}{(2k)!}\left(1+\delta\right)^{2k}.
\end{eqnarray*}
The sum of binomial coefficients arises by grouping the permutations 
$\sigma$ according to the number of registers $j$ in the set 
$\{R_1,\ldots,R_k\}$ that $\sigma$ causes to contain $\ket{\psi}$.
We therefore have
$p_{eq} \sim \sqrt{\pi k}(\frac{1+\delta}{2})^{2k}$.

We now show that the error probability cannot be less than 
$\frac{1}{4}(\frac{1+\delta}{2})^{2k}$ for the state distinguishing 
problem.%
\footnote{Note that this lower bound concerns a problem 
that is slightly more general than the problem of distinguishing 
{\em fingerprints}, because the fingerprints used in 
Section~\ref{secfingerprint} are not arbitrary but come from a known 
set of only $2^n$ states.}
Consider an optimal {\em state distinguisher\/} that acts on $k$ 
copies of $\ket{\phi}$ and $k$ copies of $\ket{\psi}$ where either 
$\ket{\phi} = \ket{\psi}$ or $|\inp{\phi}{\psi}| \le \delta$.
Let $\ket{\phi_1} = \ket{\psi_1} = \ket{0}$, and let 
$\ket{\phi_2} 
= \cos(\frac{\theta}{2})\ket{0} + \sin(\frac{\theta}{2})\ket{1}$ and 
$\ket{\psi_2} 
= \cos(\frac{\theta}{2})\ket{0} - \sin(\frac{\theta}{2})\ket{1}$, 
where $\theta = \cos^{-1}(\delta)$.
Clearly, $\ket{\phi_1} = \ket{\psi_1}$ and $\inp{\phi_2}{\psi_2} 
= \delta$.
A state distinguisher must distinguish between the state 
$\ket{a} = \ket{\phi_1}^{\otimes k} \otimes \ket{\psi_1}^{\otimes k}$ 
and the state 
$\ket{b} = \ket{\phi_2}^{\otimes k} \otimes \ket{\psi_2}^{\otimes k}$.
We now consider the probability with which a state distinguisher 
can distinguish between these states.
Since $\inp{\phi_1}{\phi_2} = \inp{\psi_1}{\psi_2} = \cos(\frac{\theta}{2})$, 
it follows that 
$\inp{a}{b} = \cos^{2k}(\frac{\theta}{2}) 
= (\frac{1+\cos \theta}{2})^{k} = (\frac{1+\delta}{2})^k$.
Now, it is known that the optimal procedure distinguishing between 
two states with inner product $\cos\alpha$ has error probability 
$\frac{1-\sin\alpha}{2} \ge \frac{1}{4}\cos^2{\alpha}$.
(This follows from an early result 
of Helstrom~\cite{helstrom:detection}, which was later strengthened
by Fuchs~\cite[Section~3.2]{fuchs:thesis}.
A clean and self-contained derivation of this result may also 
be found in~\cite{preskill:problem2}.)
Therefore, the state distinguisher must have error probability at least
$\frac{1}{4}(\frac{1+\delta}{2})^{2k}$.


\section{Errorless fingerprinting using a shared quantum key}
\label{secexactfinger}

Finally, we consider briefly the case of fingerprinting where 
Alice and Bob have a shared {\em quantum\/} key, consisting of 
$O(\log n)$ Bell states, but are required to output 
{\em classical\/} strings as fingerprints.
Is there any sense in which a quantum key can result in improved 
performance over the case of a classical key?
We observe that results in~\cite{bct:simulating} imply an improvement 
in the particular setting where the fingerprinting scheme must be exact 
(i.e., the error probability is 0) and where there is a restriction 
on the inputs that either $x=y$ or the Hamming distance between $x$ and 
$y$ is $n / 2$ (and $n$ is divisible by 4).

Under this restriction, any classical scheme with a shared key would still 
require fingerprints of length $\Omega(n)$.
On the other hand, there is a scheme with a shared quantum key of 
$O(\log n)$ Bell states that requires fingerprints of length only 
$O(\log n)$ bits.
See \cite{bct:simulating} for details (the results are partly based 
on results in \cite{BuhrmanCleveWigderson98,frankl&rodl:forbidden}).
It should be noted that if the exactness condition is relaxed to one where 
the error probability must be $O(1/n^c)$ (for a constant $c$) then there 
exists also a classical scheme with classical keys and
fingerprints of length $O(\log n)$.


\subsection*{Acknowledgments}
We thank John Preskill for references to the literature about 
optimally distinguishing between quantum states, and Andris Ambainis
for information about the origins of the classical $O(\sqrt{n})$-bit
fingerprinting scheme.
Some of this research took place while R.C. was at the CWI and while 
H.B. and R.C. were at Caltech, and the hospitality of these institutions 
is gratefully acknowledged.


\bibliographystyle{alpha}

\end{document}